\documentclass[aps,showpacs,nofootinbib,superscriptaddress,twocolumn]{revtex4}

\usepackage{amsmath}
\usepackage{graphicx}
\usepackage{float}
\usepackage{bm}
\usepackage{hyperref}
\newcommand{\WW}{{\mathrm{WW}}}

\begin{document}

\title{Double-spin asymmetry $A^{\cos\phi_S}_{LT}$ in semi-inclusive DIS at CLAS12 and EIC within the collinear framework}
\author{Xiaoyu Wang}
\affiliation{Department of Physics, Southeast University, Nanjing 211189, China}
\author{Wenjuan Mao}
\affiliation{School of Physics and Telecommunication Engineering,
Zhoukou Normal University,
Zhoukou 466000, China
}
\affiliation{Department of Physics, Southeast University, Nanjing
211189, China}
\author{Zhun Lu}
\email{zhunlu@seu.edu.cn}
\affiliation{Department of Physics, Southeast University, Nanjing 211189, China}

\begin{abstract}
We study the longitudinal-transverse double-spin asymmetry with a $\cos\phi_S$ modulation in semi-inclusive deep inelastic scattering for charged and neutral pions production. We consider the particular case in which the transverse momentum of the final state hadron is integrated out. The corresponding asymmetry may be contributed by two parts: one is the convolution of the twist-3 distribution function $g_{T}(x)$ and the unpolarized fragmentation function $D_1(z)$, the other is related to the coupling of the transversity distribution function $h_1(x)$ and the collinear twist-3 fragmentation function $\tilde{E}(z)$. We take into account both contributions to predict the $\cos\phi_S$ asymmetry at the kinematics of CLAS12 and a future Electron Ion Collider. We find that the asymmetry of pion production at CLAS12 is sizable, and $\tilde{E}(z)$ can play an important role in the asymmetry in the large-$z$ region.
\end{abstract}

\maketitle

\section{Introduction}

Understanding the partonic structure of the nucleon is one of the main tasks in QCD and hadronic physics, whereas asymmetries in semi-inclusive deep inelastic scattering (SIDIS) process with polarized beams and targets have been recognized as very useful tools for this quest.
The full description of SIDIS includes a set of transverse momentum dependent (TMD) parton distribution functions (PDFs) and fragmentation functions (FFs)~\cite{Kotzinian:1994dv,Mulders:1995dh,Bacchetta:2006tn}.
Considering the case in which the lepton beam is longitudinally polarized and the target nucleon is transversely polarized, up to twist-3 level, there are three spin or azimuthal asymmetries arising, namely, the modulations of $\cos (\phi_h-\phi_S)$, $\cos \phi_S$ and $\cos (2\phi_h-\phi_S)$, where $\phi_h$ and $\phi_S$ are the azimuthal angles of the final-state hadron and the transverse spin of the nucleon.
Among them, the $\cos (\phi_h-\phi_S)$ asymmetry is a leading twist observable contributed by the TMD PDF $g_{1T}$, and has been studied by models and experiments~\cite{Kotzinian:2006dw,Boffi:2009sh,Zhu:2011zza,Huang:2011bc}.
The other two double spin asymmetries appear in the subleading order of $1/Q$ expansion, with $Q$ the virtuality of the virtual photon.
As demonstrated in Ref.~\cite{Bacchetta:2006tn}, under the TMD framework, each asymmetry receives several contributions from the twist-3 TMD PDFs and FFs that are coupled with the twist-2 FFs and PDFs.
The roles of the twist-3 TMD PDFs on the $\cos (2\phi_h-\phi_S)$ and $\cos \phi_S$ asymmetries were studied in Ref.~\cite{Mao:2014fma} via spectator model calculations recently.

As different contributions mix together in the asymmetries at the twist-3 level,
it is difficult to disentangle individual contributions in order to access those twist-3 PDFs and FFs through SIDIS measurement.
In this work, we resort to the collinear case in which the
transverse momentum of the final-state hadron is integrated out
(or is not measured). Under this circumstance, only the
$\cos\phi_S$ asymmetry remains, because the other two
asymmetries involve $\bm k_T$-odd TMD PDFs or FFs which vanish
after the transverse momentum is integrated over. Furthermore,
the $\cos\phi_S$ asymmetry is contributed by two terms out of six:
one is the convolution of the twist-3 PDF $g_{T}^q(x)$ and the unpolarized FF $D_1^q(z)$, the other is the coupling of the transversity $h_1^q(x)$ and the collinear twist-3 chiral-odd FF $\tilde{E}^q(z)$.
Although the information of the $\bm k_T$-odd TMD PDFs and FFs is lost
in the collinear picture, there is an opportunity to focus on the remained functions that give rise to the asymmetry.
In light of this, we study the feasibility to access the twist-3 PDFs and FFs via the $\cos\phi_S$ asymmetry in double polarized SIDIS.
In particular, we will consider the effect of the FF $\tilde{E}^q(z)$, which encodes the quark-gluon-quark correlation during fragmentation.
We note that the contribution of $\tilde{E}^q(z)$ in the $\cos\phi_S$ asymmetry has not been taken into account in previous studies.
The double polarized SIDIS can be performed in the CLAS12 experiment which will soon be operational at JLab.
A future option of SIDIS is the planned Electron Ion Collider (EIC).
Thus in this paper we estimate the $\cos\phi_S$ asymmetry as functions of $x$ and $z$ at kinematics of CLAS12 and EIC.
To this end, we calculate the distribution $g_T^q(x)$ for $u$ and $d$ quarks by considering the Wandzura-Wilczek contribution~\cite{Wandzura:1977qf} and the genuine twist-3 contribution~\cite{Braun:2011aw}.
For the FF $\tilde{E}^q(z)$, we adopt an approximate relation between $\tilde{E}(z)$ and $D_1(z)$ motivated by the chiral quark model~\cite{Ji:1993qx,Yuan:2003gu}.
Furthermore, we take into account the scale dependences of the PDFs and FFs entering the description of the asymmetry.

\begin{figure*}
\centering
\includegraphics[width=0.9\columnwidth]{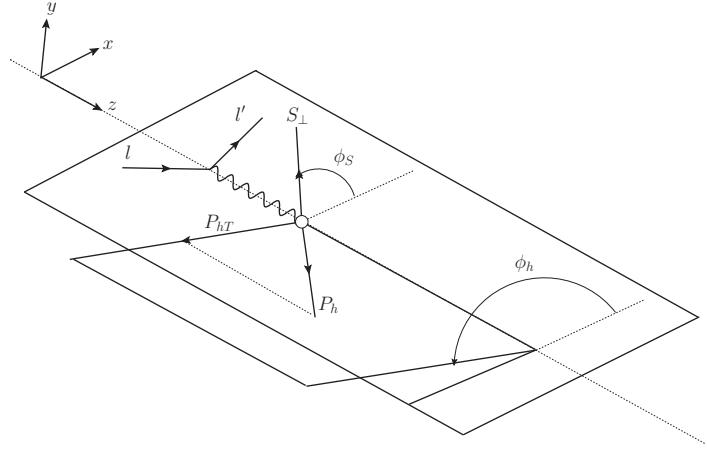}
\caption{The definition of azimuthal angles in SIDIS~\cite{Airapetian:2004tw}.
The lepton plane is defined by $l$ and $l'$. $S$ stands for the spin of the proton target, while $P_{h}$ for the momentum of the produced pion. $S_\perp$ is the transverse component of $S$ with respect to the virtual photon momentum.}
\label{lepton-hadron plane}
\end{figure*}

This paper is organized in the following way.
In Sec.~\ref{Sec.formalism}, we set up the formalism of the $\cos \phi_S$ asymmetry in SIDIS in the collinear picture.
In Sec.~\ref{Sec.numerical}, we present the numerical calculation of the asymmetries in the leptoproduction of charged and neutral pions at CLAS12 and EIC.
In Sec.~\ref{Sec.conclusion}, we summarize our work and present the conclusion.

\section{Formalism of the $\cos\phi_S$ asymmetry in SIDIS}
\label{Sec.formalism}

The process we study is the pion production semi-inclusive deep inelastic scattering using a longitudinally polarized electron beam scattered off a transversely polarized proton target:
\begin{equation}
\label{eq:sidis}
\vec{e}(\ell)+N^\uparrow(P) \longrightarrow e(\ell^\prime)+{\pi}(P_h)+X(P_X),
\end{equation}
where $l$ and $l'$ stand for the momenta of incoming and outgoing leptons, namely electron, whereas $P$ and $P_h$ denote the momenta of the target nucleon and the final-state hadron (in our case the hadron is the pion meson), respectively.
The reference frame of the process under study is shown in Fig.~\ref{lepton-hadron plane}, where the momentum of virtual photon defines the $z$ axis, in accordance with the Trento conventions~\cite{Bacchetta:2004jz}, $\phi_h$ denotes the azimuthal angle between the hadron momentum and the lepton scattering plane, while $\phi_S$ stands for the azimuthal angle of the transverse spin of the proton target.

The invariants used to express the differential cross section are defined as
\begin{align}
\label{eq:invariants}
x=\frac {Q^2}{2P \cdot q} ,\quad y=\frac{P \cdot q}{P \cdot l} ,\quad z=\frac{P \cdot P_h}{P \cdot q},\\\nonumber
 \gamma=\frac{2Mx}{Q} ,\quad Q^2=-q^2 ,\quad s=(P+l)^2.
\end{align}
As usual, $q=\ell-\ell^\prime$ is defined as the momentum of the virtual photon.
Up to twist-3 level, the six-fold ($x$, $y$, $z$, $\phi_h$, $\phi_S$ and $P_T$) double polarized differential cross section in SIDIS with a longitudinally polarized electron and a transversely polarized target has the general form~\cite{Bacchetta:2006tn}:
\begin{align}
&\frac{d^6\sigma}{dxdydzd\phi_hd\phi_SdP_{hT}^2}\,=\,
\frac{\alpha^2}{xyQ^2}\,\frac{y^2}{2(1-\varepsilon)}\,\left(1+\frac{\gamma^2}{2x}\right)\nonumber\\
&\times|\bm S_T|\lambda_e\left\{\sqrt{2\varepsilon(1-\varepsilon)}\cos \phi_S F^{\cos\phi_S}_{LT}(x,z,P_{hT})\right.\nonumber\\
&+\sqrt{2\varepsilon(1-\varepsilon)} \cos (2\phi_h-\phi_S) F^{\cos(2\phi_h-\phi_S)}_{LT}(x,z,P_{hT})\nonumber\\
& + \textrm{leading twist terms}\big{\}}\,,
\label{eq:diffcs1}
\end{align}
where $\bm S_T$ is the transverse spin vector of the nucleon, $\lambda_e$ is the helicity of the electron beam,  and $\varepsilon$ is the ratio of the longitudinal and transverse photon flux
\begin{equation}
\label{eq:epsilon}
\varepsilon=\frac{1-y-\frac{1}{4}\gamma^2y^2}{1-y+\frac{1}{2}y^2+\frac{1}{4}\gamma^2y^2}.
\end{equation}

In Eq.~(\ref{eq:diffcs1}), $F^{\cos\phi_S}_{LT}$ and $F^{\cos(2\phi_h-\phi_S)}_{LT}$ are the twist-3 structure functions that contribute to the $\cos\phi_S$ and the $\cos (2\phi_h-\phi_S)$ azimuthal asymmetries, respectively.
Particularly, $F^{\cos\phi_S}_{LT}(x,z,P_{hT})$ can be expressed as~\cite{Bacchetta:2006tn}
\begin{align}
F_{LT}^{\cos\phi_S}(x,z,P_{hT})&=\frac{2M}{Q}{\mathcal{C}}\left\{-\left(x g_T D_1 +\frac{M_h}{M} h_1\frac{\tilde{E}}{z}\right)\right.\nonumber\\
&+\frac{\bm k_T \cdot \bm p_T}{2M M_h}\left[\left(x e_T H_1^\perp-\frac{M_h}{M}g_{1T}\frac{\tilde{D^\perp}}{z}\right)\right.\nonumber\\
&\left.\left.+\left(x e^\perp_T H_1^\perp+\frac{M_h}{M}f_{1T}^\perp \frac{\tilde{G^\perp}}{z}\right)\right]\right\}\,,\label{eq:flt1}
\end{align}
where $\bm k_T$ and $\bm p_T$ are the transverse momenta of the incoming and outgoing quarks, $M_h$ and $M$ are the masses of the outgoing hadron and the target proton, and the notation $\mathcal{C}[\omega f D]$ defines the convolution:
\begin{align}
\mathcal{C}[\omega f D]&=x\sum_q e_q^2\int d^2\bm{p}_T d^2\bm{k}_T\delta^{(2)}\left(\bm k_T-\bm p_T-\bm P_{hT}/z\right)\,\nonumber \\
&\omega(\bm p_T,\bm k_T)\,f^q(x,\bm{k}_T^2)\,D^q(z,\bm{p}_{T}^2),
\end{align}
where $\omega(\bm p_T,\bm k_T)$ is an arbitrary function of $\bm p_T$ and $\bm k_T$, and the summation runs over all considered quarks and antiquarks.
In this work, we will consider the particular case in which the transverse momentum of the outgoing pion meson is integrated out, or equivalently, the case in which only the longitudinal momentum fraction $z$ of pion is measured.
Thus, after the integral $\int d^2\bm P_{hT}$ is performed, the four-fold differential cross section has the form
\begin{align}
&\frac{d^4\sigma}{dxdydzd\phi_S}=\frac{2\alpha^2}{xyQ^2}\,\frac{y^2}{2(1-\varepsilon)}\,
\left(1+\frac{\gamma^2}{2x}\right)\nonumber\\
&\times \,\sqrt{2\varepsilon(1-\varepsilon)}\,\cos \phi_S\, F^{\cos\phi_S}_{LT}\left(x,z\right)\,. \label{eq:diffcs}
\end{align}
Here, the structure function $F^{\cos\phi_S}_{LT}\left(x,z\right)$ is the collinear counterpart of the original structure function $F^{\cos\phi_S}_{LT}\left(x,z,P_{hT}\right)$~\cite{Bacchetta:2006tn}
\begin{align}
F^{\cos\phi_S}_{LT}\left(x,z\right) & = \int d^2 \bm{P}_{hT} F^{\cos\phi_S}_{LT}\left(x,z,P_{hT}\right) \nonumber\\
& =  -x\sum_q e_q^2\frac{2M}{Q}\bigg{(}xg_T^q(x)D_1^q(z)\nonumber\\
&+\frac{M_h}{M}h^q_1(x)\frac{\tilde E^q(z)}{z}\bigg{)}\,.\label{eq:fltcol}
\end{align}
Eq.~(\ref{eq:fltcol}) contains the convolution of the twist-3 distribution $g_T^q(x)$ and the twist-2 FF $D_1^q(z)$, as well as that of the twist-3 fragmentation function $\tilde{E}^q(z)$ and the twist-2 PDF $h_1^q(x)$.

\begin{figure*}
\centering
\includegraphics[width=1.6\columnwidth]{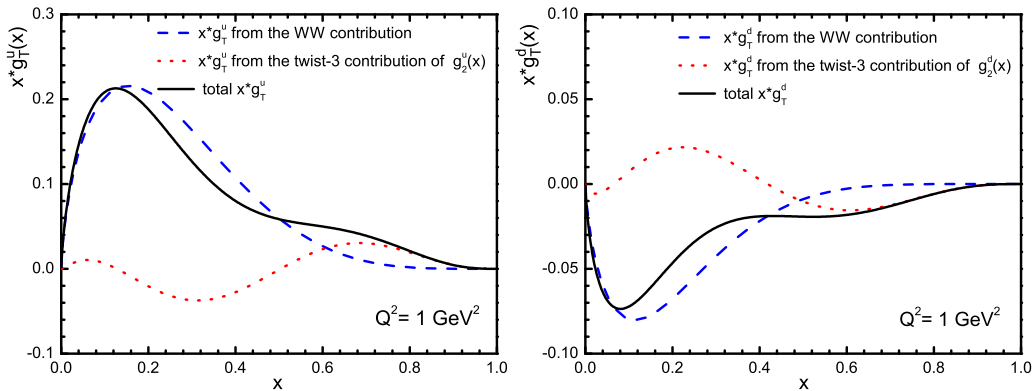}
\caption{The twist-3 distribution function $xg_T^q(x)$ for up and down quarks vs $x$ at  $Q^2\ = 1\rm{GeV}^2$.}
\label{fig:gt}
\end{figure*}

The longitudinal-transverse spin asymmetry may be defined as
\begin{align}
A_{LT} \sim {\sigma(+\lambda_e, \bm S_T )-\sigma(-\lambda_e, \bm S_T )
\over \sigma(+\lambda_e, \bm S_T )+\sigma(-\lambda_e, \bm S_T )},
\end{align}
which is consistent with the notation of previous experimental measurements\cite{Huang:2011bc}.
Thus, the $x$-dependent $\cos \phi_S$ asymmetry can be defined as
\begin{align}
\label{eq:autx}
&A_{LT}^{\cos\phi_S}(x) \nonumber \\
=&\frac{\int dy\int dz\frac{\alpha^2}{xyQ^2}\frac{y^2}{2(1-\varepsilon)}(1+\frac{\gamma^2}{2x})
\sqrt{2\varepsilon(1-\varepsilon)}F^{\cos\phi_s}_{LT}(x,z)}
{\int dy\int dz\frac{\alpha^2}{xyQ^2}\frac{y^2}{2(1-\varepsilon)}(1+\frac{\gamma^2}{2x})F_{UU}(x,z)}\,,
\end{align}
where $F_{UU}$ is the unpolarized structure function:
\begin{equation}
F_{UU}(x,z)=x\sum_q e_q^2f_1^q(x)D_1^q(z)\,,\label{eq:fuucol}
\end{equation}
with $f_1^q(x)$ and $D_1^q(z)$ being the unpolarized PDF and FF, respectively.
In a similar way, the $\cos \phi_S$ asymmetry as a function of $z$ can be written as
\begin{align}
\label{eq:autz}
&A_{LT}^{\cos\phi_s}(z) \nonumber\\
&=\frac{\int dx\int dy\frac{\alpha^2}{xyQ^2}\frac{y^2}{2(1-\varepsilon)}(1+\frac{\gamma^2}{2x})
\sqrt{2\varepsilon(1-\varepsilon)}F^{\cos\phi_s}_{LT}(x,z)}
{\int dx\int dy\frac{\alpha^2}{xyQ^2}\frac{y^2}{2(1-\varepsilon)}(1+\frac{\gamma^2}{2x})F_{UU}(x,z)}.
\end{align}


The twist-3 distribution function $g_T^q(x)$ can be expressed as the combination of the spin-dependent structure functions $g_1(x)$ and $g_2(x)$~\cite{Barone:2001sp}:
\begin{equation}
\frac{1}{2}\sum_q \,e_q^2\,g^q_T(x)=g_1(x)+g_2(x)\,, \label{eq:gt12}
\end{equation}
where $g_1(x)$ is the leading twist structure function contributed from the helicity PDFs
\begin{align}
&g_1(x)=\frac{1}{2}\sum_q e_q^2 g^q_1(x)\,,
\end{align}
and $g_2(x)$ is the structure function related to the transverse spin of the target proton, which can be separated into two parts
\begin{equation}
g_2(x)=g^{\WW}_2(x)+g^{\rm{tw-3}}_2(x)\,. \label{eq:g2}
\end{equation}
Here $g^{\WW}_2(x)$ and $g^{\rm{tw-3}}_2(x)$ are the Wandzura-Wilczek and genuine twist-3 contributions to $g_2(x)$, respectively.

In the absence of the higher twist contribution $g^{\rm{tw-3}}_2(x)$, the structure function $g_2(x)$ is determined by the structure function $g_1(x)$
\begin{equation}
g_2  \; \stackrel{\WW}{\approx} g^{\WW}_2(x)= -g_1(x)+\int _x^1 dy\frac{g_1(y)}{y},
\label{eq:gtww}
\end{equation}
which is usually referred to as the Wandzura-Wilczek approximation~\cite{Wandzura:1977qf}.
A number of theoretical~\cite{Gockeler:2000ja,Gockeler:2005vw,Balla:1997hf,Ball:1996tb,Blumlein:1996vs,Blumlein:1998nv,Kivel:2000rb,
Radyushkin:2000ap,Anikin:2001ge,Teryaev:1995um,Metz:2008ib,Accardi:2009au,Braun:2011aw} and experimental~\cite{Adams:1994id,
Abe:1998wq,Anthony:2002hy,Amarian:2003jy,Zheng:2004ce,Kramer:2005qe} works have been carried out to study the validity of this approximation.
Particularly, in Ref.~\cite{Braun:2011aw} a result of $g^{\rm{tw-3}}_2(x)$ for proton and neutron target obtained from the convolution integrals of the light-cone wave functions was presented at the reference scale $Q^2=1\ \mathrm{GeV}^2$ ($\bar{x}=1-x$):
\begin{align}
g^{\rm{tw-3}}_{2,p}(x)&=0.0436772(\mathrm{ln}x+\bar{x}+\frac{1}{2}\bar{x}^2)+\bar{x}^3(1.57357\nonumber\\
&-5.94918\bar{x}+6.74412\bar{x}^2-2.19114\bar{x}^3),\label{eq:gtw31}\\
g^{\rm{tw-3}}_{2,n}(x)&=0.0655158(\mathrm{ln}x+\bar{x}+\frac{1}{2}\bar{x}^2)+\bar{x}^3(0.130996 \nonumber\\
&-1.12101\bar{x}+2.31342\bar{x}^2-1.20598\bar{x}^3),
\label{eq:gtw32}
\end{align}
which is used to compare with the SLAC and JLab data.

In this work we apply the results in Eqs.~(\ref{eq:gtw31}) and (\ref{eq:gtw32}) to obtain the twist-3 PDF $g_T^q(x)$.
To do this, we combine Eqs.~(\ref{eq:gt12}), (\ref{eq:g2}) and (\ref{eq:gtww}) to yield
\begin{align}
\frac{1}{2}g^q_T(x)&=\frac{1}{2}\int _x^1 dy\frac{g_1^{q}(y)}{y}+g^{\rm{tw-3},q}_2(x),
\end{align}
where $g^{\rm{tw-3},q}_2(x)$ is the contribution to $g^{\rm{tw-3}}_2(x)$ from $q$ flavor.
In the following we assume that $g^{\rm{tw-3}}_2(x)$ is mainly contributed by $u$ and $d$ quarks, which should be valid in the valence region.
After applying the isospin symmetry
\begin{align}\nonumber
g^{\rm{tw-3}}_{2,p}(x)=\frac{4}{9}g^{\rm{tw-3},u}_2(x)+\frac{1}{9}g^{\rm{tw-3},d}_2(x),\\
g^{\rm{tw-3}}_{2,n}(x)=\frac{1}{9}g^{\rm{tw-3},u}_2(x)+\frac{4}{9}g^{\rm{tw-3},d}_2(x),
\end{align}
we can obtain the expression for $g^q_T(x)$
\begin{align}
g^u_T(x)&=\int _x^1 dy\frac{g^u_1(y)}{y}+\frac{6}{5}(4g^{\rm{tw-3}}_{2,p}(x)-g^{\rm{tw-3}}_{2,n}(x)),\label{eq:gut}\\
g^d_T(x)&=\int _x^1 dy\frac{g^d_1(y)}{y}+\frac{6}{5}(4g^{\rm{tw-3}}_{2,n}(x)-g^{\rm{tw-3}}_{2,p}(x)).\label{eq:gdt}
\end{align}

\begin{figure*}
\centering
\includegraphics[width=1.6\columnwidth]{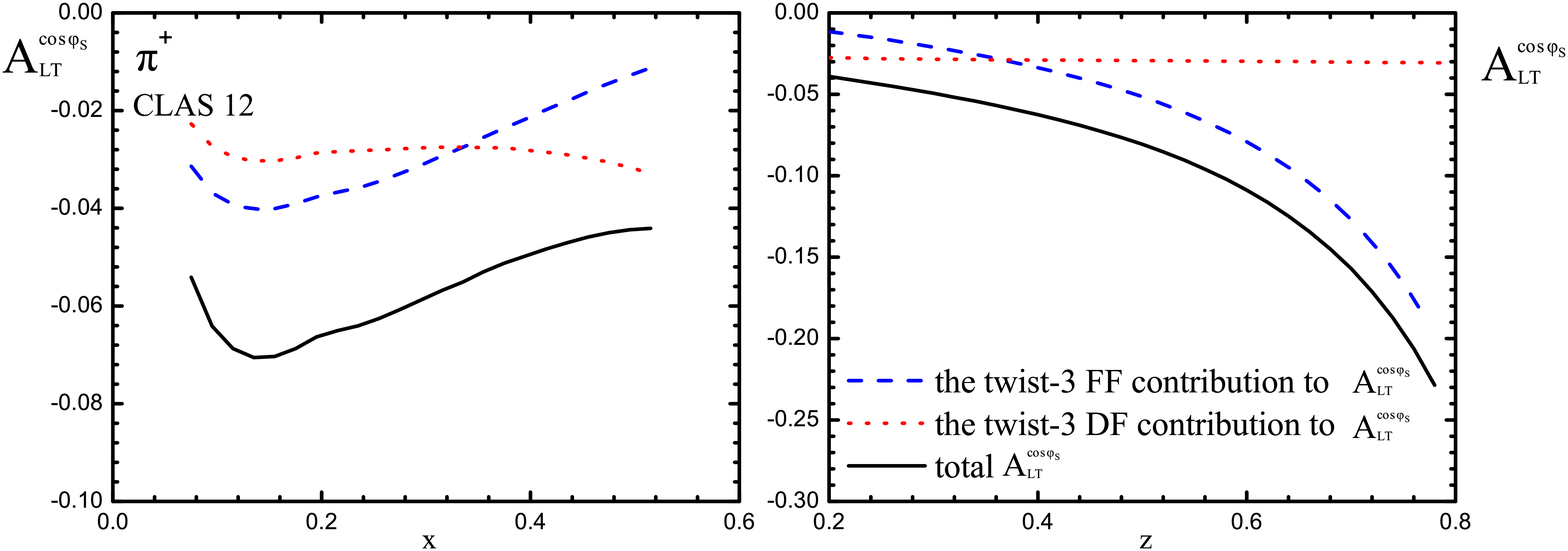}
\includegraphics[width=1.6\columnwidth]{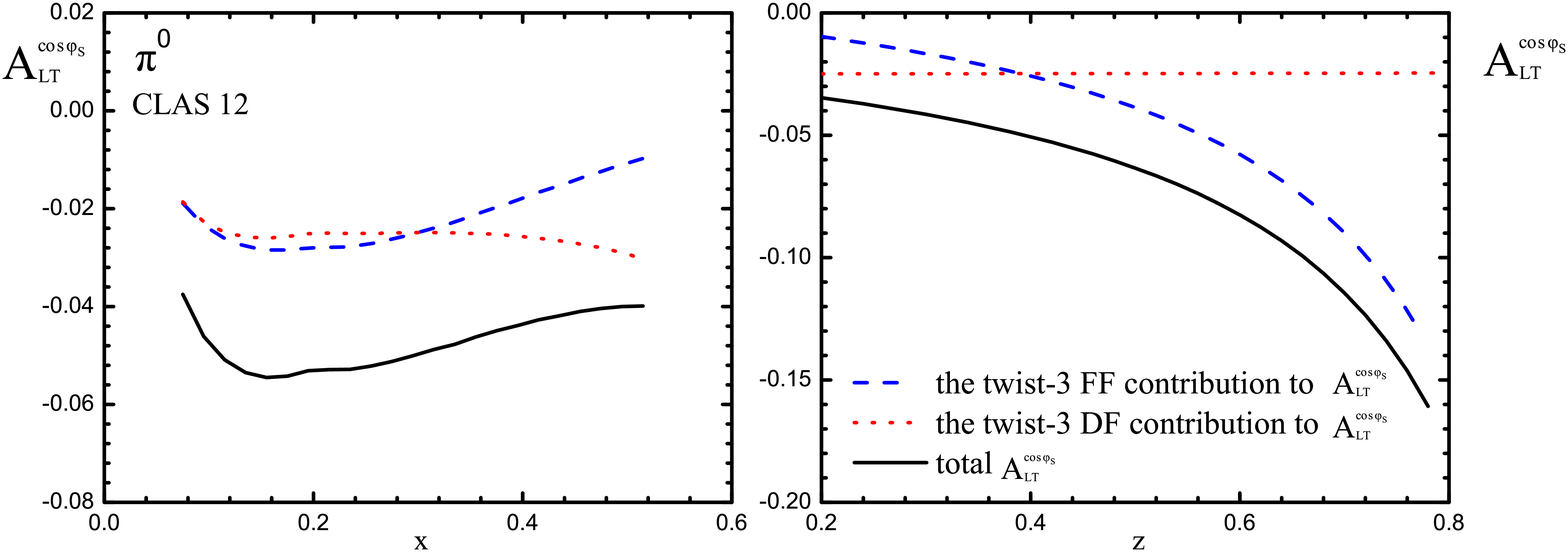}
\includegraphics[width=1.6\columnwidth]{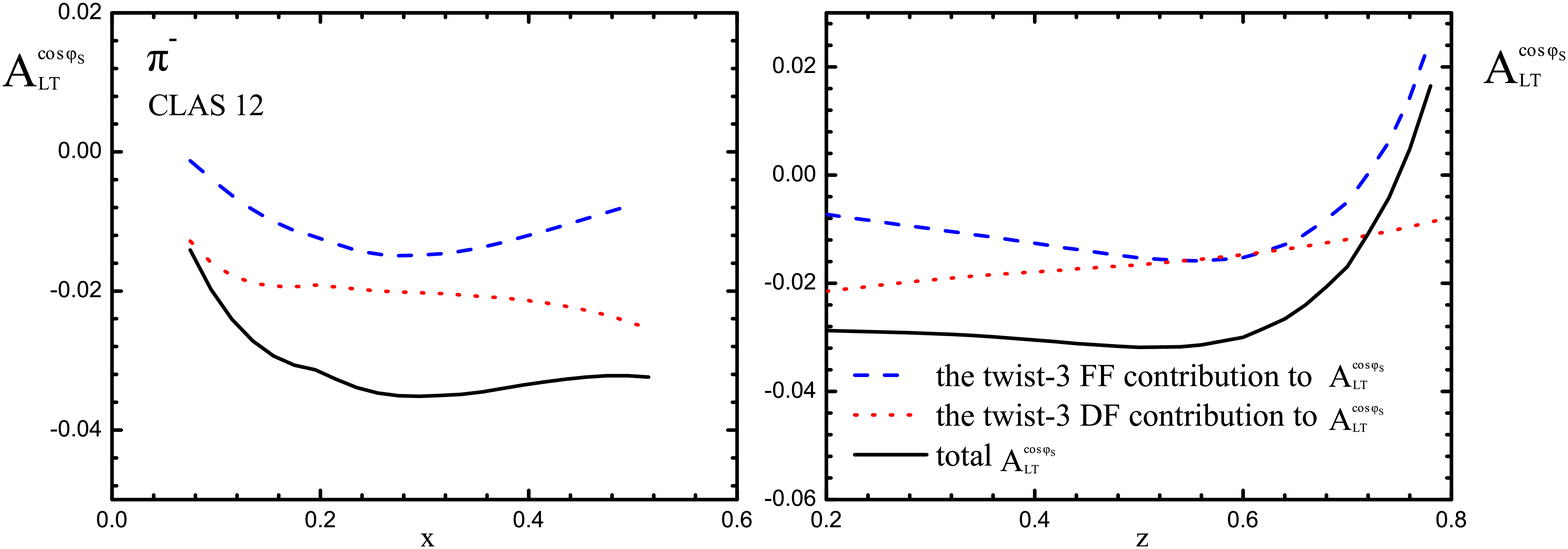}
\caption{Longitudinal-transverse double-spin asymmetry $A^{\cos \phi_S}_{LT}$ of $\pi^+$, $\pi^0$ and $\pi^-$ production in SIDIS at CLAS12. The left panels show the $x$-dependent asymmetry, while the right ones the $z$-dependent asymmetry.}
\label{fig:clas12}
\end{figure*}

To estimate the asymmetry $A_{LT}^{\cos\phi_S}$, we also need the knowledge of the
chiral-odd twist-3 FF $\tilde{E}^q(z)$, which involves the quark-gluon-quark correlation.
Currently there is no theoretical and experimental information on $\tilde{E}^q(z)$.
The only constraint on $\tilde{E}^q(z)$ is the equation of motion relation~\cite{Mulders:1995dh}
\begin{align}
{E^q(z)\over z} = {\tilde{E}^q(z)\over z} + {m_q\over M_h} D^q_1(z), \label{eq:eom}
\end{align}
where $E^q(z)$ is the twist-3 FF~\cite{Jaffe:1993xb} encoded in the quark-quark correlation during fragmentation, $m_q$ is the current quark mass, and $M_h$ is the mass of final state hadron.
$E^q(z)$ was studied by the chiral quark model~\cite{Yuan:2003gu} and the spectator model~\cite{Gamberg:2003pz}.
The effect of $E(z)$ on the beam SSA $A_{LU}^{\sin\phi_h}$ has also been calculated~\cite{Yuan:2003gu,Gamberg:2003pz}.
To estimate the DSA contributed by $\tilde{E}^q(z)$, we apply Eq.~(\ref{eq:eom}) and adopt the chiral quark model result for $E^q(z)$~\cite{Yuan:2003gu}~\footnote{In Ref.~\cite{Yuan:2003gu}, the definition of the twist-3 FF, denoted by $\hat e^q(z)$, is slightly different from the definition of $E^q(z)$ in Ref.~\cite{Bacchetta:2006tn}. However, they are related by $M \hat e^q(z) = M_h E^q(z)$ according to Eq.~(5) in Ref.~\cite{Yuan:2003gu}.}
\begin{equation}
E^q(z) = \frac{m_q^\prime}{M_h}\frac{z}{1-z}D_1^q(z), \label{eq:ez}
\end{equation}
with $m_q^\prime$ the constituent quark mass.
In principle the current quark mass should be much smaller than the constituent quark mass and $m_h$, so that the second term on the r.h.s of Eq.~(\ref{eq:eom}) should be negligible.
As a rough estimate, we assume that values of the two masses are the same for simplicity.
This assumption may be crude.
However, as we will show in the next section, the assumption will not change the main result of our paper.
Thus, in our estimation $\tilde{E}^q(z)$ is proportional to the unpolarized FF $D_1^q(z)$
\begin{equation}
\tilde E^q(z) = \frac{m_q^\prime}{M_h}\frac{z^2}{1-z}D_1^q(z).\label{eq:etilde}
\end{equation}
For the quark mass we choose $m_q^\prime\approx M/3$, following the choice in Ref.~\cite{Yuan:2003gu}.

As for the transversity $h_1(x)$ in Eq.~(\ref{eq:fltcol}), we adopt the standard parametrization from Ref.~\cite{Anselmino:2013vqa} (at the initial scale $Q^2=2.41~\textrm{GeV}^2$)£º
\begin{equation}
h_1^q(x)=\frac{1}{2}\mathcal{N}_q^T(x)[f_{1}^q(x)+g_1^q(x)]\,, \label{eq:h1}
\end{equation}
with
\begin{equation}
\label{eq:n}
 \mathcal{N}_q^T(x)=N_q^T\,x^\alpha(1-x)^\beta
 \frac{(\alpha+\beta)^{\alpha+\beta}}{\alpha^\alpha\beta^\beta}.
\end{equation}
In order to be in consistence with the choices in Ref.~\cite{Anselmino:2013vqa},  we apply the GRV98 leading-order (LO) parametrization~\cite{Gluck:1998xa} for the unpolarized PDF $f_1^q(x)$.
For the helicity PDF $g_1^q(x)$ appearing in Eqs.~(\ref{eq:gut}), (\ref{eq:gdt}) and (\ref{eq:h1}), we adopt the GRSV2000 LO parametrization~\cite{Gluck:2000dy}.
For $D_1^q(z)$ appearing in Eq.~(\ref{eq:fltcol}), (\ref{eq:fuucol}) and (\ref{eq:etilde}), we choose the LO set of the DSS parametrization~\cite{deFlorian:2007aj}.

\begin{figure*}
\centering
\includegraphics[width=1.6\columnwidth]{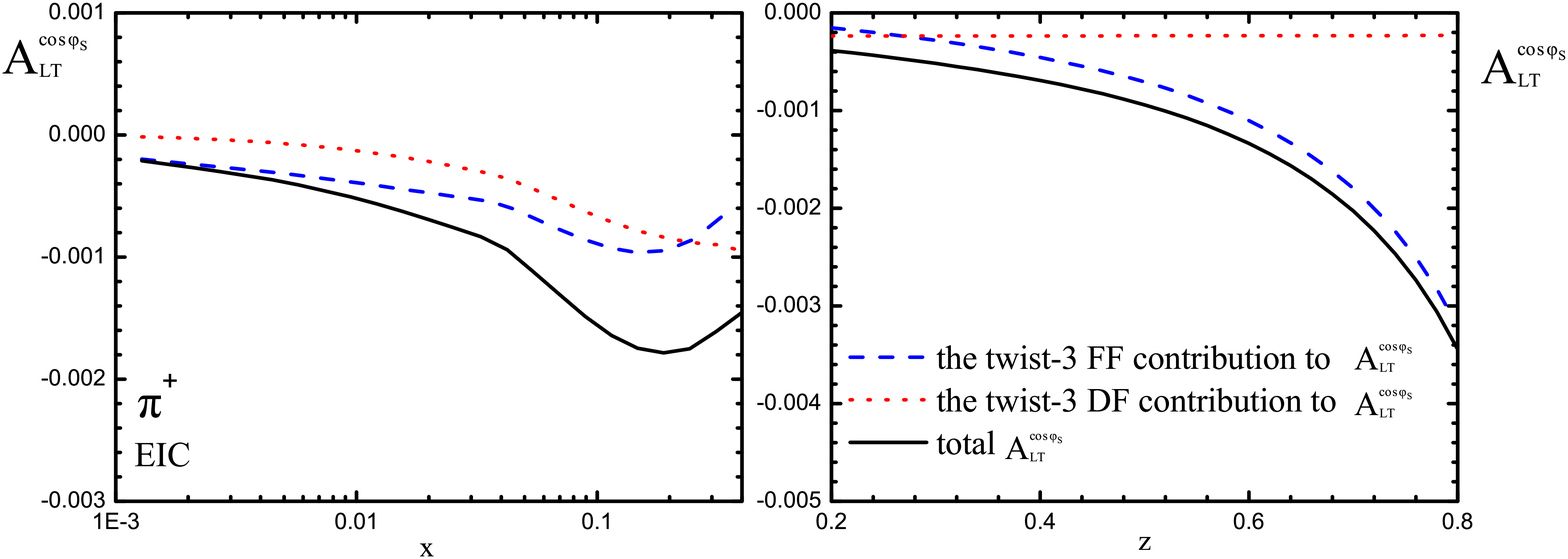}
\includegraphics[width=1.6\columnwidth]{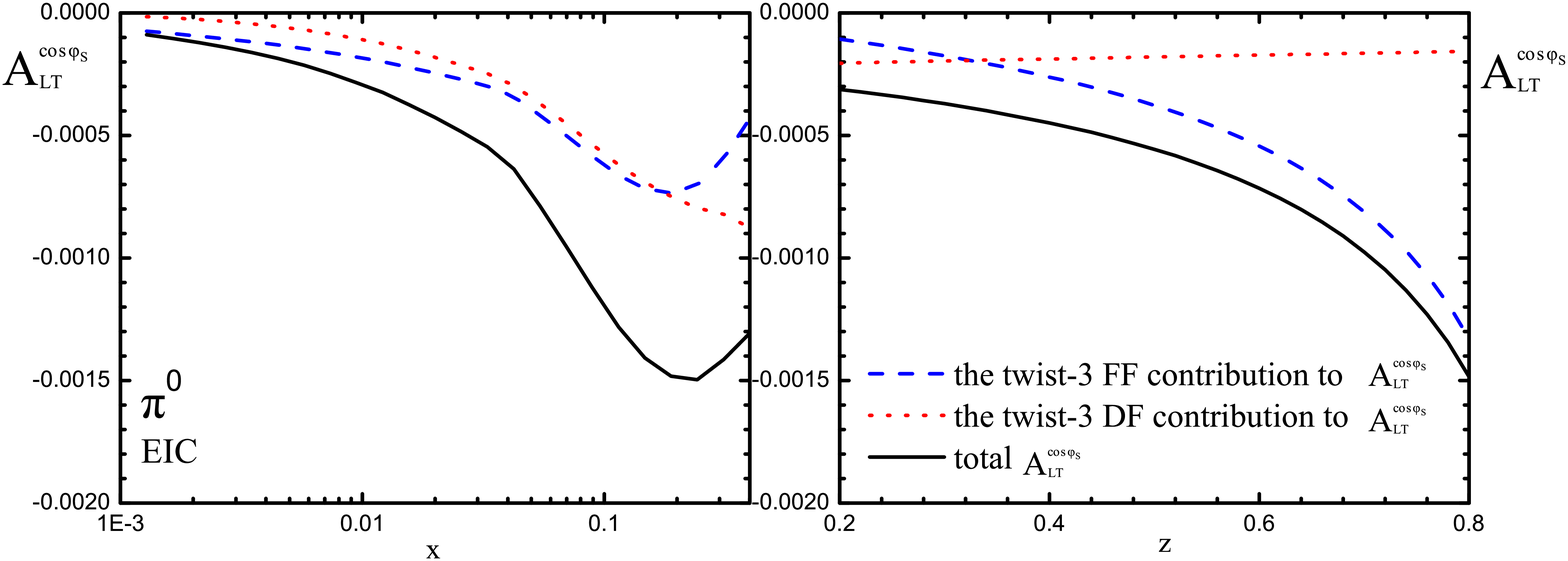}
\includegraphics[width=1.6\columnwidth]{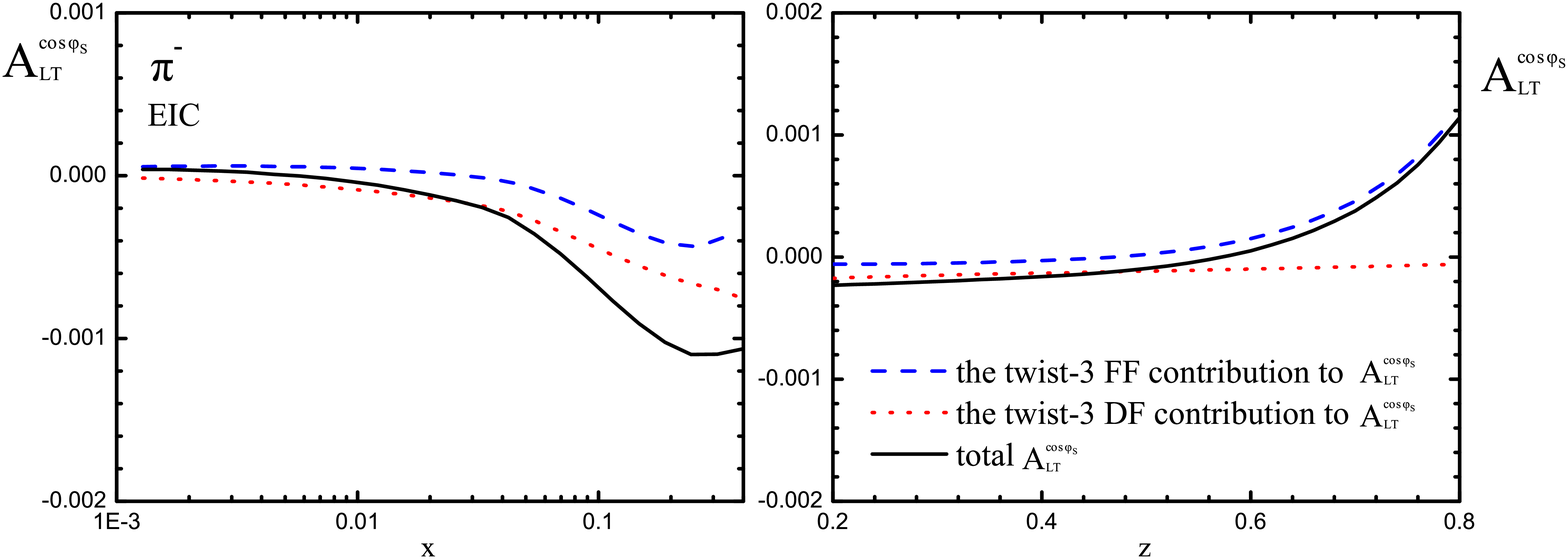}
\caption{Similar to Fig.~\ref{fig:clas12}, but at EIC for $\sqrt{s}=45\, \mathrm{GeV}$.}
\label{fig:eic}
\end{figure*}

\section{Numerical Estimate}
\label{Sec.numerical}

In this section, we perform the numerical calculation to obtain an estimate of the $\cos\phi_S$ asymmetry by utilizing the formalism presented in Sec.~\ref{Sec.formalism}.
In our calculation we will take into account the kinematical configurations at CLAS12 and EIC.

In Fig.~\ref{fig:gt}, we plot the twist-3 PDF $xg_T^q(x)$ vs $x$ at the reference scale $Q^2\ =1\ \rm{GeV}^2$.
The left and right panels denote the results for $u$ quark and $d$ quark, respectively. The dashed lines show the results calculated from the Wandzura-Wilczek relation in Eq.~(\ref{eq:gtww}), the dotted lines show the contribution from the twist-3 part of $g_2(x)$, and the solid lines denote the total $xg_T(x)$. From the curves, we can draw a conclusion that $g_T^u(x)$ is positive, while $g_T^d(x)$ is negative, and both the $g^{\rm{tw-3}}_2(x)$ contribution are sizable. The size of $g_T^d(x)$ is somewhat smaller than that of $g_T^u(x)$.

Since the PDFs and FFs given in Eqs. (\ref{eq:gut}), (\ref{eq:gdt}), (\ref{eq:etilde}) and (\ref{eq:h1}) are given at certain fixed scales, the evolution to other scales is necessary.
The scale dependence of $g_T(x)$ is determined from that of $g^{q,WW}_T(x)$ and $g^{q,\rm{tw-3}}_T(x)$:
 \begin{align}
 \label{eq:gtqq2}
 g^q_T(x,Q^2)&=g^{q,\WW}_T(x,Q^2)+g^{q,\rm{tw-3}}_T(x,Q^2)\,.
 \end{align}
In this work we assume that the $Q^2$ dependence of $g^{q,WW}_T(x,Q^2)$ comes from that of $g_1^q(x,Q^2)$ for simplicity
\begin{equation}
 g^{q,\WW}_T(x,Q^2)= \int _x^1 dy\frac{g_1^q(y,Q^2)}{y}\,.
\label{eq:gtwwevo}
\end{equation}
To evolve the twist-3 contribution $g_T^{\rm{tw-3}}$, we adopt the non-singlet evolution kernel for $g^{\rm{tw-3}}_2(x)$
\begin{align}
P^{NS,z\rightarrow1}_{q,F}(z)=2C_F\left[\frac{1}{(1-z)_+}+\frac{3}{4}\delta(1-z)\right]-{1\over 2}N_C\delta(1-z).
\label{eq:split}
\end{align}
The above kernel is a simpler version of the exact evolution based on the large-$N_c$ and large-$x$ approximation.
The same evolution was also used in Refs.~\cite{Braun:2009mi,Braun:2011aw}.
As shown in Ref.~\cite{Braun:2011aw}, the scale dependence of the twist-3 contribution $g^{\rm{tw-3}}_2(x)$ calculated from Eq.~(\ref{eq:split}) almost coincides with the result from exact evolution.
Therefore, in this paper we apply Eq.~(\ref{eq:split}) for the evolution of $g^{\rm{tw-3}}_2(x)$ for simplicity.
To perform numerics we implement Eq.~(\ref{eq:split}) in the \sc HOPPET~\cite{Salam:2008qg} \rm package.
The HOPPET package is also applied to evolve the transversity $h_1(x)$ after including chiral-odd LO splitting functions in the code.

The FF $E(z)$ used in our estimation is obtained at the chiral symmetry breaking scale.
At higher scale the relation (\ref{eq:ez}) might breakdown because the evolutions of $E(z)$ and $D_1(z)$ are different.
However, as a rough estimate, we will assume that the $Q^2$ dependence of ${E}(z)$, as well as  $\tilde{E}(z)$,  is the same as that of $D_1(z)$.

The kinematical configuration used to calculate the ${\cos\phi_S}$ asymmetry at CLAS12 is as follows~\cite{Matevosyan:2015gwa},
\begin{align}
&0.072< x < 0.532,\quad 0.2< z <0.8, E_e=11\ \mathrm{GeV}, \nonumber \\
& W^2 >\ 4\ \mathrm{GeV}^2,\quad 1 < Q^2 <6.3\ \mathrm{GeV}^2,
\end{align}
where $W$ is the invariant mass of the photon-nucleon system:
$$W^2=(P+q)^2 \approx \frac{1-x}{x}Q^2.$$
In the upper, central and lower panels of Fig.~\ref{fig:clas12}, we show the $\cos \phi_S$ asymmetries at CLAS12 for $\pi^+$, $\pi^0$ and $\pi^-$, respectively.
In each panel, we plot the $x$-dependent (left figure) and $z$-dependent (right figure) asymmetries.
The dashed and the dotted lines correspond to the asymmetries contributed by the twist-3 PDF $g_T^q(x)$ and the twist-3 FF $\tilde{E}^q(z)$, respectively, while the solid lines depict the sums of the two contributions.

\begin{figure}
\centering
\scalebox{0.175}{\includegraphics*[100pt,0pt][1490pt,540pt]{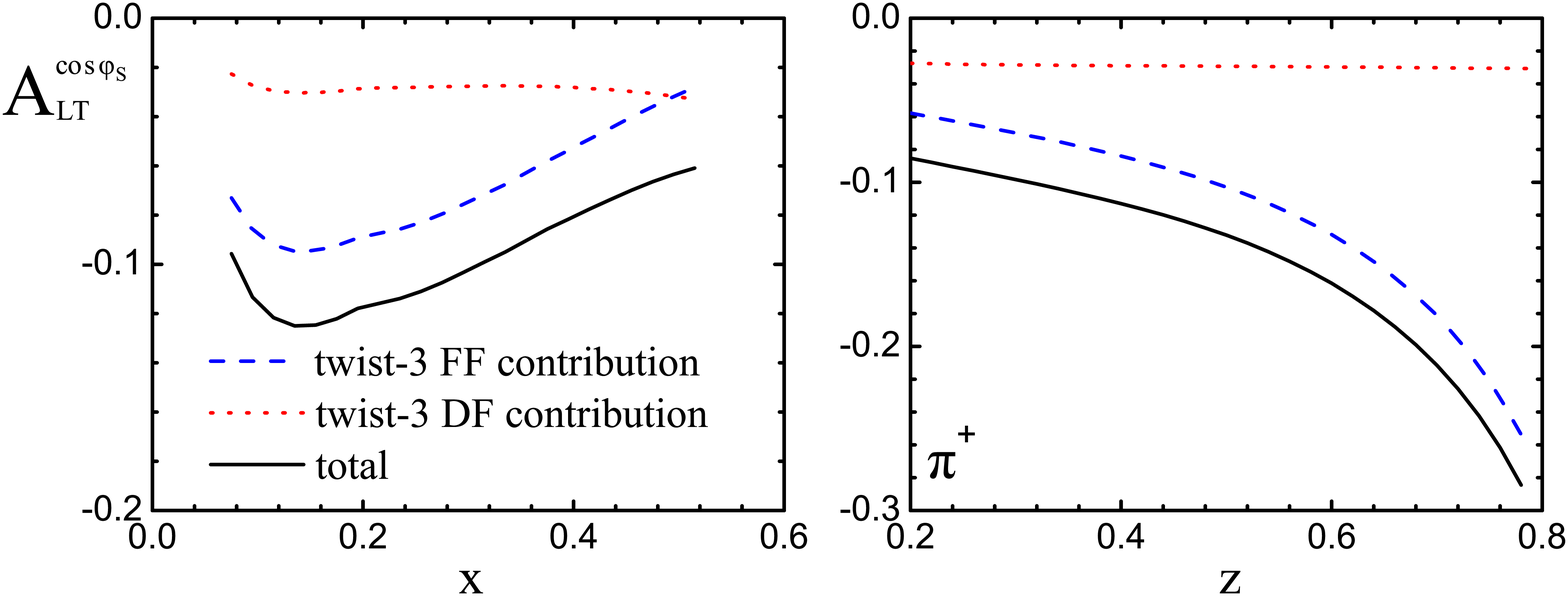}}
\scalebox{0.175}{\includegraphics*[100pt,0pt][1490pt,540pt]{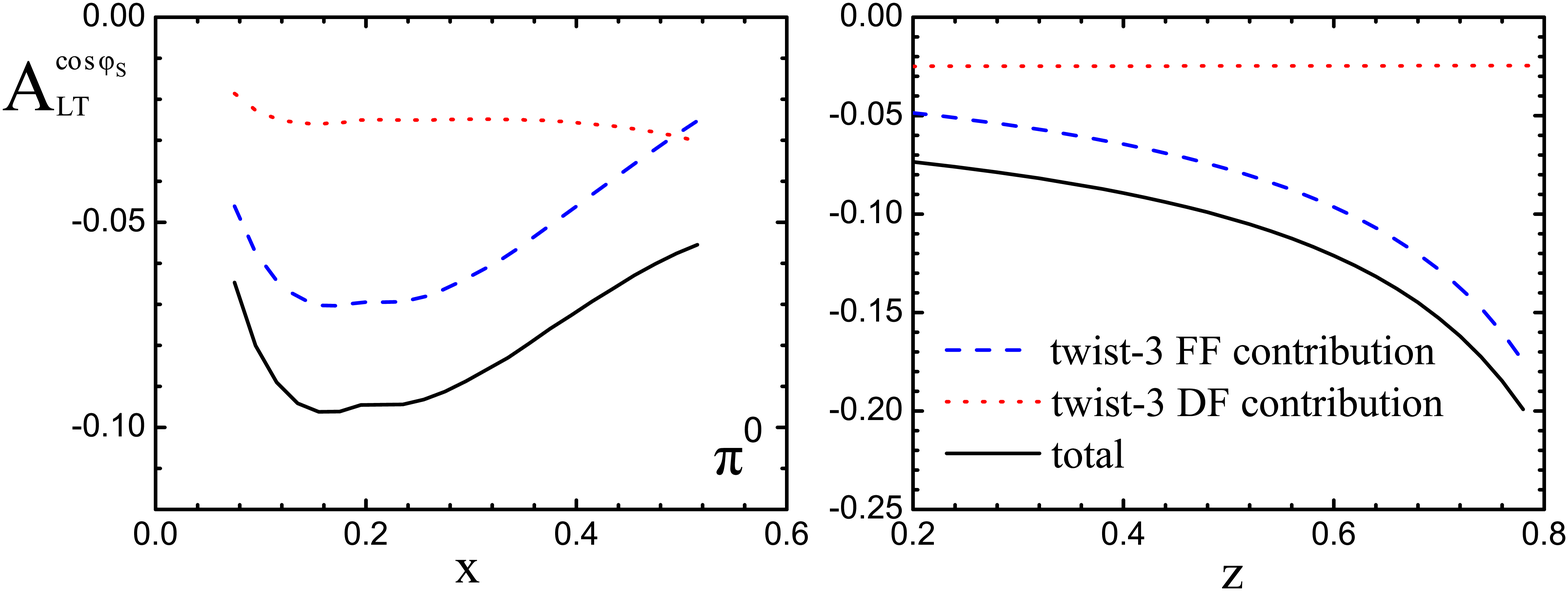}}
\scalebox{0.175}{\includegraphics*[100pt,0pt][1490pt,540pt]{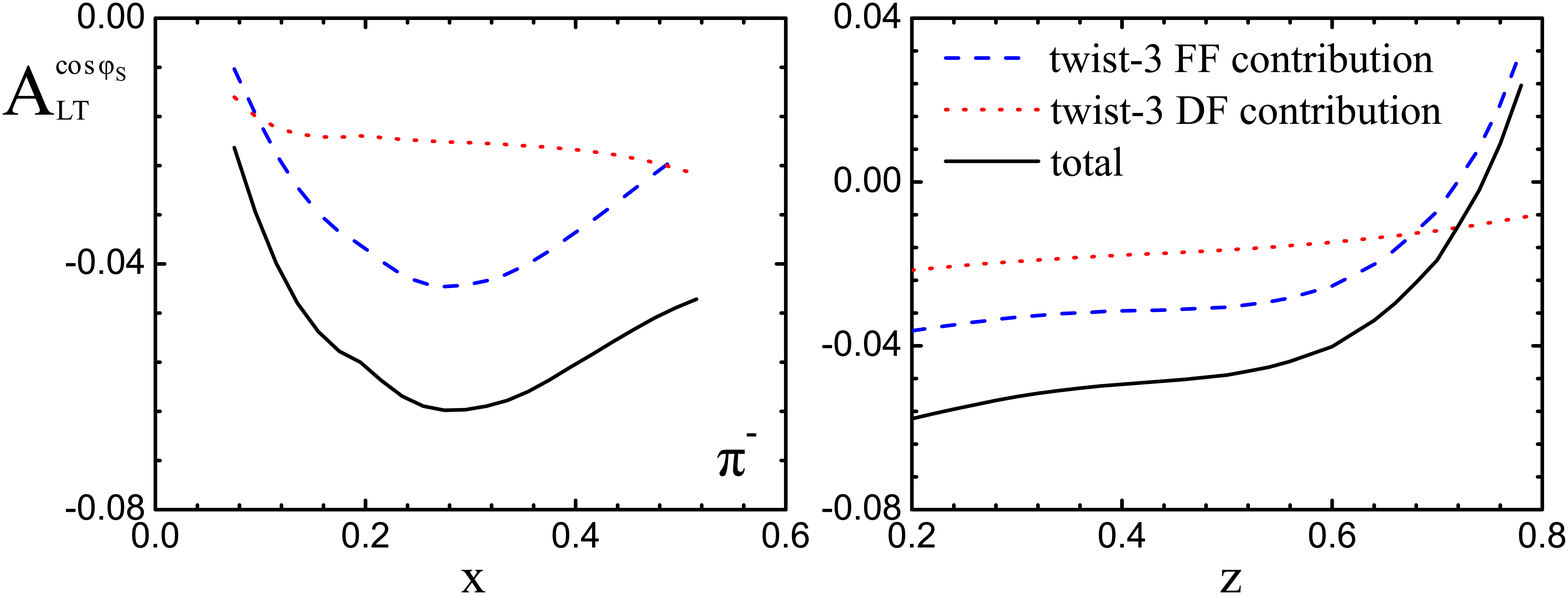}}
\caption{Longitudinal-transverse double-spin asymmetry $A^{\cos \phi_S}_{LT}$ of $\pi^+$, $\pi^0$ and $\pi^-$ production in SIDIS at CLAS12, but using Eq.~(\ref{eq:etilde2}) for $\tilde{E}$ instead of Eq.~(\ref{eq:etilde}).}
\label{fig:clas122}
\end{figure}

We find the $\cos\phi_S$ asymmetries for both the charged and neutral pions are sizable at CLAS12, about several percent.
Another observation is that the size of the asymmetries for $\pi^+$ and $\pi^0$  is larger than the size of the asymmetry for $\pi^-$ production.
Therefore, it is feasible to measure the $\cos\phi_S$ asymmetry through the CLAS12 experiments, in the case the transverse momentum of the final hadron is not measured.
From the $x$-dependent curves one can see that the asymmetries for charged and neutral pions are all negative.
This is because both the $\tilde{E}^q(z)$ contribution and the $g_T^q(x)$ contribution to the $x$-dependent asymmetries are negative.
We also find that the size of the contribution from $\tilde{E}^q(z)$ is comparable to that from $g_T^q(x)$.
In addition, the magnitude of the $x$-dependent asymmetries increases with $x$ in the small-$x$ region, reaches a peak around $0.1<x<0.2$, then it decreases mildly as $x$ increases for both charged and neutral pion production.
The $z$-dependent asymmetries for different pions are mostly negative, except the asymmetry for $\pi^-$ in the large-$z$ region.
The contribution of $g_T^q(x)$ has a weak dependence on $z$, especially in the case of $\pi^+$ and $\pi^-$.
However, the magnitude of the $\tilde{E}^q(z)$ contribution increases rapidly with increasing $z$, since there is a factor $z^2/(1-z)$ in the expression of $\tilde{E}^q(z)$.
In the large-$z$ region, the contribution from $\tilde{E}^q(z)$ might dominate over that from $g_T^q(x)$.
Hence, there is an opportunity to access the $h_1(x)\otimes\tilde{E}^q(z)$ term, provided that the  statistic of the data in the large-$z$ region is substantial.


To calculate the $\cos\phi_S$ asymmetry at EIC, we adopt the following kinematical cuts ~\cite{Matevosyan:2015gwa}
\begin{align}
&Q^2>1 \mathrm{GeV}^2, \quad 0.001<x<0.4,\quad 0.01<y<0.95,\nonumber\\
&0.2<z<0.8,\quad \sqrt{s}=45\ \mathrm{GeV},\quad W>5\ \mathrm{GeV}.
\end{align}
In Fig.~\ref{fig:eic}, we plot $A_{LT}^{\cos\phi_S}$ of charged and neutral pions vs $x$ and $z$ at EIC, similar to the format in Fig.~\ref{fig:clas12}.
Although the sign and the shape of the asymmetries for different pion productoin at EIC are similar to those at CLAS12, it is found that the $\cos\phi_S$ asymmetry at the kinematical configuration of EIC is much smaller (less than 0.3\%).
This is because the asymmetry we study is at the twist-3 level, at which the effect will be suppressed by $1/Q$, and the averaged $Q$ value at EIC is much higher than that at CLAS12.


We end this section with two comments.
Firstly, to obtain the collinear results in Eqs.~(\ref{eq:diffcs}) and (\ref{eq:fltcol}) from the fully differential cross section in Eq.~(\ref{eq:diffcs1}), we have performed a formal, exact analytical integration over $P_{hT}$ in the range $[0,+\infty]$.
As shown in Ref.~\cite{Boglione:2011wm}, in parton-model based approaches
with on mass-shell partons, there are kinematical constraints on the maximum size of the parton transverse momentum:
\begin{equation}
 \begin{cases}
k_{T}^2\leq(2-x)(1-x)Q^2, ~~~\textrm{for}~~0< x< 1 
, \\
k_{T}^2\leq \frac{x(1-x)} {(1-2x)^2}\, Q^2, ~~~~~~~~~~~~\textrm{for}~~x< 0.5,
\end{cases}\label{constraints}
 \end{equation}
and hence of $P_{hT}$.
In literature these constraints are higher-twist kinematical effects and are often neglected
for leading-twist observables.
The study in Ref.~\cite{Boglione:2011wm} showed that, for higher-twist observables (such as the twist-3 Cahn effect), applying the additional requirement in Eq.~(\ref{constraints}) will lead to
different results with respect to the usual phenomenological approach based on analytical integration over an unlimited range of $\bm k_T$ values.
To check if the kinematical constraints like those given in Eq.~(\ref{constraints}) is relevant
in the study of the asymmetry $A_{LT}^{\cos\phi_S}$ contributed by $g_T$ and $\tilde{E}$,
we apply the constraints in Eq.~(\ref{constraints}) to perform the integration over $\bm P_{hT}$, $\bm k_T$ and $\bm p_T$ numerically for the first line in Eq.~(\ref{eq:flt1}).
For more details, we assume the Gaussian form for the transverse momentum dependence of the PDFs and FFs.
In this case we find no difference with respect to the collinear results calculated from Eq.~(\ref{eq:fltcol}).
Furthermore, we verify that the results do not change when we vary the Gaussian widths of the PDFs and FFs.
This is different from the Cahn effect for which the kinematical constraints lead to a different result from the analytical integration~\cite{Boglione:2011wm}.
The reason may come from that fact that there is a $k_T$-dependent prefactor $\bm k_T\cdot \bm{P}_{hT}/|\bm{P}_{hT}| $ associated with the Cahn effect, while in our case such prefactor does not appear in the convolution $g_T\otimes D_1$  or $h_1\otimes \tilde{E}$.

Secondly, as mentioned in the previous section, in our calculation we have assumed that the current quark mass is the same as the constituent quark mass.
As a check, we also consider the case in which the current quark mass term in Eq.~(\ref{eq:eom}) is neglected.
The fragmentation $\tilde{E}(z)$ thus has the form:
\begin{align}
\tilde{E}(z) \approx E (z) =  \frac{m_q^\prime}{M_h}\frac{z}{1-z}D_1^q(z) . \label{eq:etilde2}
\end{align}
We apply Eq.~(\ref{eq:etilde2}) to recalculate the asymmetry $A_{LT}^{\cos\phi_S}$ at CLAS12 and plot the result in Fig.~\ref{fig:clas122}.
We find that in this case the asymmetry contributed by the fragmentation function $\tilde{E}$ is more significant, as a factor of $z$ is removed in Eq.~(\ref{eq:etilde}).
Specifically, the $x$-dependent asymmetry from $\tilde{E}$ is almost doubled in the intermediate $x$ region, and the asymmetry at smaller $z$ region is enhanced.
However, the shape of the total symmetry is generally similar to the result in the previous calculation: i.e., the $x$-dependent asymmetry peak at around $x\sim 0.2$, and the magnitude of the $z$-dependent asymmetry increases with increasing $z$.
Hence, the calculation according to Eq.~(\ref{eq:etilde2}) shows that there is still a good opportunity to access $\tilde{E}$ through measuring the asymmetry $A_{LT}^{\cos\phi_S}$ at CLAS12.

\section{Conclusion}
\label{Sec.conclusion}

In this work, we have studied the $\cos\phi_S$ asymmetry in double polarized SIDIS.
Particularly, we have focused on the case that the transverse momentum of the final-state hadron is integrated out.
Under this circumstance the asymmetry arises from two contributions, namely, the convolution of $g_T^q(x)$ and $D_1^q(z)$, as well as that of  $h_1^q(x)$ and $\tilde{E}^q(z)$.
We have included both contributions to estimate the $\cos\phi_S$ asymmetry for charged and neutral pions at the kinematics of CLAS12 and EIC.
To do this we have gone beyond the Wandzura-Wilczek approximation and adopted an analysis of $g_2^{\textrm{tw-3}}(x)$ to obtain the genuine twist-3 part of $g_T^q(x)$.
Furthermore, motivated by the chiral quark model, we have employed an approximate relation between the twist-3 FF $\tilde{E}^q(z)$ and the unpolarized FF $D_1^q(z)$.
In addition, we have considered the evolution effect of the twist-2 and twist-3 PDFs and FFs in the calculation.
The numerical prediction shows that the asymmetries for the charged and neutral pions are all sizable at CLAS12, about several percent.
In contrast, the asymmetries at EIC are much smaller due to the suppression in the large-$Q$ region.
Although for the $x$-dependent asymmetry the size of the contribution from $\tilde{E}^q(z)$ is comparable to that from $g_T^q(x)$, we find that the asymmetry in the large-$z$ region is completely dominated by the convolution of $h_1^q(x)$ and $\tilde{E}^q(z)$.
Therefore, it might be promising to access the unknown twist-3 FF $\tilde{E}^q(z)$ via the measurement of the $\cos \phi_S$ asymmetry of pion production in SIDIS with the collinear picture.

\section*{Acknowledgements}
This work is partially supported by the National Natural Science
Foundation of China (Grants No.~11575043, and No.~11120101004), and
by the Qing Lan Project.


\begin{thebibliography}{99}

\bibitem{Kotzinian:1994dv}
  A.~Kotzinian,
  Nucl.\ Phys.\ B {\bf 441}, 234 (1995).

\bibitem{Mulders:1995dh}
  P.~J.~Mulders and R.~D.~Tangerman,
  Nucl.\ Phys.\ B {\bf 461}, 197 (1996)
  Erratum: [Nucl.\ Phys.\ B {\bf 484}, 538 (1997)]
  [hep-ph/9510301].

\bibitem{Bacchetta:2006tn}
  A.~Bacchetta, M.~Diehl, K.~Goeke, A.~Metz, P.~J.~Mulders and M.~Schlegel,
  JHEP {\bf 0702}, 093 (2007).

\bibitem{Kotzinian:2006dw}
  A.~Kotzinian, B.~Parsamyan and A.~Prokudin,
  Phys.\ Rev.\ D {\bf 73}, 114017 (2006).

\bibitem{Boffi:2009sh}
  S.~Boffi, A.~V.~Efremov, B.~Pasquini and P.~Schweitzer,
  Phys.\ Rev.\ D {\bf 79}, 094012 (2009).

\bibitem{Zhu:2011zza}
  J.~Zhu and B.~Q.~Ma,
  Phys.\ Lett.\ B {\bf 696}, 246 (2011).

\bibitem{Huang:2011bc}
  J.~Huang {\it et al.} [Jefferson Lab Hall A Collaboration],
  doi:10.1103/PhysRevLett.108.052001
  [arXiv:1108.0489 [nucl-ex]].

\bibitem{Mao:2014fma}
  W.~Mao, Z.~Lu, B.~Q.~Ma and I.~Schmidt,
  Phys.\ Rev.\ D {\bf 91}, no. 3, 034029 (2015).

\bibitem{Wandzura:1977qf}
  S.~Wandzura and F.~Wilczek,
  Phys.\ Lett.\ B {\bf 72}, 195 (1977).

\bibitem{Braun:2011aw}
  V.~M.~Braun, T.~Lautenschlager, A.~N.~Manashov and B.~Pirnay,
  Phys.\ Rev.\ D {\bf 83}, 094023 (2011).

  \bibitem{Ji:1993qx}
  X.~D.~Ji and Z.~K.~Zhu,
  hep-ph/9402303.

\bibitem{Yuan:2003gu}
  F.~Yuan,
  Phys.\ Lett.\ B {\bf 589}, 28 (2004).

\bibitem{Airapetian:2004tw}
  A.~Airapetian {\it et al.} (HERMES Collaboration),
  Phys.\ Rev.\ Lett.\  {\bf 94}, 012002 (2005).

\bibitem{Bacchetta:2004jz}
  A.~Bacchetta, U.~D'Alesio, M.~Diehl and C.~A.~Miller,
  Phys.\ Rev.\ D {\bf 70}, 117504 (2004).

\bibitem{Barone:2001sp}
  V.~Barone, A.~Drago and P.~G.~Ratcliffe,
  Phys.\ Rept.\  {\bf 359}, 1 (2002).

\bibitem{Gockeler:2000ja}
  M.~Gockeler, R.~Horsley, W.~Kurzinger, H.~Oelrich, D.~Pleiter, P.~E.~L.~Rakow, A.~Schafer and G.~Schierholz,
  Phys.\ Rev.\  D {\bf 63} (2001) 074506
  [arXiv:hep-lat/0011091].

\bibitem{Gockeler:2005vw}
  M.~Gockeler, R.~Horsley, D.~Pleiter, P.~E.~L.~Rakow, A.~Schafer, G.~Schierholz, H.~Stuben and J.~M.~Zanotti,
  Phys.\ Rev.\ D {\bf 72}, 054507 (2005)
  [arXiv:hep-lat/0506017].

\bibitem{Balla:1997hf}
  J.~Balla, M.~V.~Polyakov and C.~Weiss,
  Nucl.\ Phys.\  B {\bf 510} (1998) 327
  [arXiv:hep-ph/9707515].

\bibitem{Ball:1996tb}
  P.~Ball and V.~M.~Braun,
  Phys.\ Rev.\  D {\bf 54}, 2182 (1996)
  [arXiv:hep-ph/9602323].

\bibitem{Blumlein:1996vs}
  J.~Bl\"umlein and N.~Kochelev,
  Nucl.\ Phys.\  B {\bf 498} (1997) 285
  [arXiv:hep-ph/9612318].

\bibitem{Blumlein:1998nv}
  J.~Bl\"umlein and A.~Tkabladze,
  Nucl.\ Phys.\  B {\bf 553} (1999) 427
  [arXiv:hep-ph/9812478].

\bibitem{Kivel:2000rb}
  N.~Kivel, M.~V.~Polyakov, A.~Sch\"afer and O.~V.~Teryaev,
  Phys.\ Lett.\  B {\bf 497}, 73 (2001)
  [arXiv:hep-ph/0007315].

\bibitem{Radyushkin:2000ap}
  A.~V.~Radyushkin and C.~Weiss,
  Phys.\ Rev.\  D {\bf 63}, 114012 (2001)
  [arXiv:hep-ph/0010296].

\bibitem{Anikin:2001ge}
  I.~V.~Anikin and O.~V.~Teryaev,
  Phys.\ Lett.\  B {\bf 509}, 95 (2001)
  [arXiv:hep-ph/0102209].

\bibitem{Teryaev:1995um}
  O.~V.~Teryaev,
  in the Proceedings of
 ``Prospects of Spin Physics at HERA, Zeuthen, Germany, 28-31 Aug 1995,''
  Eds.~J.~Bl\"umlein and W.-D.~Nowak, Hamburg, Germany, 1995
  (DESY-95-200), pp.~132-142 [arXiv:hep-ph/0102296].



\bibitem{Metz:2008ib}
  A.~Metz, P.~Schweitzer and T.~Teckentrup,
  Phys.\ Lett.\ B {\bf 680}, 141 (2009)
  [arXiv:0810.5212 [hep-ph]].

\bibitem{Accardi:2009au}
  A.~Accardi, A.~Bacchetta, W.~Melnitchouk and M.~Schlegel,
  JHEP {\bf 0911}, 093 (2009)
  [arXiv:0907.2942 [hep-ph]].

\bibitem{Adams:1994id}
D. Adams et al. [Spin Muon Collaboration (SMC)],
Phys. Lett. B 336 (1994) 125 [arXiv:hepex/9408001].

\bibitem{Abe:1998wq}
K. Abe et al. [E143 collaboration],
Phys. Rev. D 58 (1998) 112003 [arXiv:hep-ph/9802357].

\bibitem{Anthony:2002hy}
  P.~L.~Anthony {\it et al.} [E155 Collaboration],
  Phys.\ Lett.\ B {\bf 553}, 18 (2003)
  [hep-ex/0204028].

\bibitem{Amarian:2003jy}
M. Amarian et al. [Jefferson Lab E94-010 Collaboration], Phys. Rev. Lett. 92 (2004) 022301
[arXiv:hep-ex/0310003].

\bibitem{Zheng:2004ce}
X. Zheng et al. [Jefferson Lab Hall A Collaboration], Phys. Rev. C 70 (2004) 065207
[arXiv:nucl-ex/0405006].

\bibitem{Kramer:2005qe}
[8] K. Kramer et al., Phys. Rev. Lett. 95 (2005) 142002 [arXiv:nucl-ex/0506005].

\bibitem{Jaffe:1993xb}
  R.~L.~Jaffe and X.~D.~Ji,
  Phys.\ Rev.\ Lett.\  {\bf 71}, 2547 (1993)
  [hep-ph/9307329].

\bibitem{Gamberg:2003pz}
  L.~P.~Gamberg, D.~S.~Hwang and K.~A.~Oganessyan,
  Phys.\ Lett.\ B {\bf 584}, 276 (2004)
  [hep-ph/0311221].

  \bibitem{Anselmino:2013vqa}
  M.~Anselmino, M.~Boglione, U.~D'Alesio, S.~Melis, F.~Murgia and A.~Prokudin,
  Phys.\ Rev.\ D {\bf 87}, 094019 (2013).

\bibitem{Gluck:1998xa}
  M.~Gluck, E.~Reya and A.~Vogt,
  Eur.\ Phys.\ J.\ C {\bf 5}, 461 (1998).

\bibitem{Gluck:2000dy}
  M.~Gluck, E.~Reya, M.~Stratmann and W.~Vogelsang,
  Phys.\ Rev.\ D {\bf 63}, 094005 (2001).

\bibitem{deFlorian:2007aj}
  D.~de Florian, R.~Sassot and M.~Stratmann,
  Phys.\ Rev.\ D {\bf 75}, 114010 (2007).

\bibitem{Braun:2009mi}
  V.~M.~Braun, A.~N.~Manashov and B.~Pirnay,
  Phys.\ Rev.\ D {\bf 80}, 114002 (2009).

\bibitem{Salam:2008qg}
  G.~P.~Salam and J.~Rojo,
  Comput.\ Phys.\ Commun.\  {\bf 180}, 120 (2009).

\bibitem{Matevosyan:2015gwa}
  H.~H.~Matevosyan, A.~Kotzinian, E.~C.~Aschenauer, H.~Avakian and A.~W.~Thomas,
  Phys.\ Rev.\ D {\bf 92}, no. 5, 054028 (2015).


\bibitem{Boglione:2011wm}
  M.~Boglione, S.~Melis and A.~Prokudin,
  Phys.\ Rev.\ D {\bf 84}, 034033 (2011)
  doi:10.1103/PhysRevD.84.034033
  [arXiv:1106.6177 [hep-ph]].

\end{thebibliography}
\end{document}